\documentstyle[preprint,floats,aps,prb,epsf]{revtex}	
\tightenlines
\draft
\preprint{submitted to {\em Physical Review B} \hskip189pt 
LA-UR-00-2950}							
\title{Muon Spin Relaxation Study of (La, Ca)MnO$_3$}
\author{R. H. Heffner and J. E. Sonier}
\address{MS K764, Los Alamos National Laboratory, Los Alamos, New Mexico 87545}
\author{D. E. MacLaughlin}
\address{Department of Physics, University of California, Riverside, California 
92521-0413}
\author{G. J. Nieuwenhuys}
\address{Kamerlingh Onnes Laboratorium, Leiden University, 2300 RA Leiden, The 
Netherlands}
\author{G. M. Luke}
\address{McMaster University, Hamilton, Ontario, L8P 4N1 Canada}
\author{Y. J. Uemura}
\address{Department of Physics, Columbia University, New York, New York 10027}
\author{William Ratcliff II and S-W. Cheong}
\address{Department of Physics and Astronomy, Rutgers University, Piscataway, NJ 08854}
\author{G. Balakrishnan}
\address{University of Warwick, Coventry CV4 7AL, UK}

\date{Draft date \today}							

\begin{document}	\maketitle					
\newpage \begin{abstract}					
We report predominantly zero field muon spin relaxation measurements in a series of Ca-doped LaMnO$_3$ compounds which includes the colossal magnetoresistive manganites. Our principal result is a systematic study of the spin-lattice relaxation rates $1/T_1$ and magnetic order parameters in the series La$_{1-x}$Ca$_x$MnO$_3, x = 0.0, 0.06, 0.18, 0.33, 0.67$ and $1.0$. In LaMnO$_3$ and CaMnO$_3$ we find  very narrow critical regions near the N\'{e}el temperatures $T_N$ and  temperature independent $1/T_1$ values above T$_N$.  From the $1/T_1$ in LaMnO$_3$ we derive an exchange integral $J = 0.83$ meV which is consistent with the mean field expression for $T_N$.  All of the doped manganites except CaMnO$_3$ display anomalously slow, spatially inhomogeneous spin-lattice relaxation below their ordering temperatures.  In the ferromagnetic (FM) insulating La$_{0.82}$Ca$_{0.18}$MnO$_3$ and ferromagnetic conducting La$_{0.67}$Ca$_{0.33}$MnO$_3$ systems we show that there exists a bi-modal distribution of $\mu$SR rates $\lambda_f$ and $\lambda_s$ associated with relatively \lq fast\rq~ and \lq slow\rq~ Mn fluctuation rates, respectively.  A physical picture is hypothesized for these FM phases in which the fast Mn rates  are due to overdamped spin waves characteristic of a disordered FM, and the  slower Mn relaxation rates  derive from distinct, relatively insulating regions in the sample.  Finally, likely muon sites are identified, and evidence for muon diffusion in these materials is discussed.
\end{abstract}							
\vspace{12pt} 


\section{Introduction}
The strong coupling of the spin, charge and lattice degrees of freedom in the colossal magnetoresistive (CMR) manganites gives rise to a wide variation in ground-state behavior, including paramagnetism, ferro-and antiferromagnetism, charge ordering (CO),  and insulating and conducting charge transport \cite{manganites}.  The earliest studies of these systems \cite{Jonker50,Wollen55} involved the substitution of a divalent alkaline earth (R = Sr, Ca, Ba) for trivalent La in the perovskite structured LaMnO$_3$, an insulating antiferromagnet.  This substitution  produces a hole at a Mn site (i.e., Mn$^{3+}$ changes to Mn$^{4+}$) which can hop between Mn ions with different valence states via the ferromagnetic double exchange (DE) interaction \cite{Zener}.   Linkage between the DE interation and charge mobility provided an early explanation of the paramagnetic/insulating to ferromagnetic/conducting transition  in La$_{1-x}$R$_x$MnO$_3$.  More recent experiments \cite{tokura} and subsequent theoretical analyses \cite{millis} have also established the importance of the local Jahn-Teller (JT) structural perturbation in the Mn$^{3+}$ state.    Thus, both the  JT and DE interactions are  important in determining the ground state properties and magnetic ordering temperatures in these materials.  
The JT and DE interactions are also thought to induce the formation of magnetoelastic polarons, local structural distortions surrounded by polarized Mn spins, in CMR compounds.  The degree of polaron formation depends upon the degree of cation size mismatch between the La atom and the alkaline earth atom.\cite{catsize}  Sr produces the least distortion and Ba the most.  These concepts comprise the fundamental ideas behind the strong coupling of spin, charge and lattice degrees of freedom in CMR materials.  In this paper we study the spin dynamics of the La$_{1-x}$Ca$_x$MnO$_3$ system using the muon spin relaxation ($\mu$SR) technique,\cite{musr}  thus providing a valuable complement to other experiments probing the charge and lattice degrees of freedom in the these materials.

The La$_{1-x}$Ca$_x$MnO$_3$ system has been extensively studied because it exhibits a wide variety of ground states as a function of $x$ below room temperature. \cite{phasediagram}  At temperatures below about $700$ K the system is orthorhombic for all Ca concentrations.  The end members LaMnO$_3$ and CaMnO$_3$ are both antiferromagnetic (AFM) insulators with different magnetic structures. The addition of Ca to LaMnO$_3$ produces a canted AFM insulating state for $0 < x \lesssim 0.07$, a ferromagnetic (FM) insulating state (followed by CO at lower temperatures) for $0.07 < x \lesssim 0.21$, and a FM metallic state for $0.21 < x \lesssim 0.50$.  Above $x = 0.50$ the system remains insulating with a high temperature CO transition observed for $0.50 \leq x \lesssim 0.875$,  followed by AFM at lower temperatures.  Experimentally, transport measurements \cite{trnspt} are consistent with the formation of magnetoelastic polarons in the paramagnetic state of La$_{1-x}$Ca$_x$MnO$_3$, and both neutron scattering \cite{pdf} and x-ray fine structure analysis \cite{xafs} find evidence for the persistence of JT distortions significantly below the FM transition.  Neutron scattering has been used to study the low-temperature magnetic properties in FM perovskites, \cite{dispersion} and has revealed a broad peak centered around zero energy transfer which coexists with Mn spin waves near $T_C$. \cite{lynn}

\subsection{$\mu$SR experiments and data analysis}
The $\mu$SR technique employed here involves the intersititial implantation of positive muons with essentially 100\% spin polarization, oriented antiparallel to the initial muon momentum vector.  The $\mu^+$ decays with a mean lifetime $\tau_\mu$ of $2.2$~$\mu$s into a positron and two undetected neutrinos.  The positron is emitted preferentially along the muon spin direction; detection of the positron emission rate in counters aligned along the initial muon spin direction allows one to monitor the time rate of decay of the muon spin polarization along its initial spin direction. The positron emission rate $dN/dt$ is given by 
\begin{equation}
dN/dt = B + N_0 (1/\tau_\mu) \exp(-t/\tau_{\mu})[1 \pm A G_z(t)],
\label{eq:dndt}
\end{equation}
where $B$ is a background (usually independent of time), $N_0$ is a normalization factor, $A$ is the average asymmetry of the decay angular distribution (typically $0.2 - 0.3$), and $G_z(t)$ describes the time rate of decay of the muon spin polarization.  The $+$ or $-$ signs are appropriate for positron counters in the direction of, or opposite to, the muon spin, respectively.  

The $\mu$SR data presented here were taken at the M20 muon channel at TRIUMF (Vancouver, Canada) and on the GPS spectrometer at the Paul Scherrer Institute  (Villigen, Switzerland).  These data could be fit to a relaxation function $G_z(t)= G_{\rm osc}(t) + G_{\rm rlx}(t)$,  corresponding to oscillating and relaxing terms, respectively.  In zero applied field the oscillating component $G_{\rm osc}(t)$ occurs in a magnetically ordered state and is given by 

\begin{equation}
G_{\rm osc}(t) = \Sigma_i (A_{\rm osc})_i\exp(-t/T_{2i}) \cos(2\pi\nu_{\mu i} t + \phi_{\mu i}),
\label{eq:Gosc} 
\end{equation}
where $(A_{\rm osc})_i$ is the amplitude of the $i^{th}$ precessing component, $\phi_{\mu i}$ is a phase angle and $\nu_{\mu i}$ and $1/T_{2i}$ are the corresponding muon precession frequency and inhomogeneous damping rate. We define $\Sigma_i (A_{\rm osc})_i \equiv A_{\rm osc}$.  The existence of multiple components can be due to more that one muon lattice site and/or different microscopic magnetic environments (even for a single muon lattice site).  The quantity $2\pi\nu_{\mu i} = \gamma_\mu B_i$, with $\gamma_\mu$ the muon gyromagnetic ratio ($8.54 \times 10^4$ Hz/Oe) and $B_i$ the local magnetic field.  

In our experiments the relaxing component $G_{\rm rlx}(t)$  frequently approximates a stretched-exponential form:

\begin{equation}
G_{\rm rlx}(t) = A_{\rm rlx} \exp[-(t/T_1)^K], 
\label{eq:strexp}
\end{equation}
where $1/T_1$ is a characteristic spin-lattice relaxation rate.  The  polycrystalline averages of $A_{\rm rlx}$ and $A_{\rm osc}$ in a magnetically ordered state (where both $A_{\rm rlx}$ and $A_{\rm osc}$ are non-zero) are $A_{\rm rlx} = 1/3$ and $A_{\rm osc} = 2/3$.  For rapid fluctuations the muon relaxation rate $1/T_1$ is given by $1/T_1 \propto \gamma^2_{\mu} \sum_q |\delta B(q)|^2 \tau(q)$, where $|\delta B(q)|$ is the amplitude of the fluctuating local field and $\tau(q)$ is the Mn-ion correlation time.  The sum over the characteristic momenta $q$ of the Mn-ion excitations occurs because the muon is a local probe.  When the exponent $K < 1$ in Eq. (3) a distribution of $1/T_1$ values is implied,\cite{Istratov99} and thus $|\delta B(q)|$ and/or $\tau(q)$ are distributed. Typically, near a  magnetic phase transition the correlation time becomes longer, causing $1/T_1$ to increase, a phenomenon known as critical slowing down.  Below the ordering temperature, the fluctuating amplitude decreases and $1/T_1$ is reduced.

The stretched exponential form in Eq. (\ref{eq:strexp}) is often  useful to parameterize the data because it involves only a few parameters and can be used to approximate $G_{\rm rlx}(t)$ over a wide temperature range.  However, for the materials under investigation here, we have found that for large $1/T_1$ and small $K$  the relaxation function can often be better approximated over a limited temperature range by the sum of two exponentials:  

\begin{equation}
G_{\rm rlx}(t) = A_f \exp(-\lambda_f t) + A_s \exp(-\lambda_s t), 
\label{eq:twoexp} 
\end{equation}

where $\lambda_f$ and $\lambda_s$ correspond to fast and slow local-field fluctuation rates $\tau^{-1}$, respectively.  This fitting function is for a bi-modal distribution of fluctuation rates, and its appropriateness is discussed in detail in Section III below.

\subsection{Sample preparation}
The La$_{1-x}$Ca$_x$MnO$_3$ samples studied here were polycrystalline materials prepared either as boules in an optical floating zone furnace ($0 \leq x \leq 0.18$) or as sintered powders ($0.33 \leq x \leq 1.0$).  In general, polycrystalline materials of (La,Ca)MnO$_3$ are more compositionally homogeneous than comparable volumes of single crystals, because of relatively less Ca evaporation and more stable growth conditions.   The samples were characterized using x-ray diffraction, resistivity and susceptibility.  All samples were greater than 98\% single phase.

\section{Experimental results: An Overview}
\subsection{LaMnO$_3$ and CaMnO$_3$: Antiferromagnets}
The compounds LaMnO$_3$ and CaMnO$_3$ are the AFM end members of the (La,Ca)MnO$_3$ series.  They are both insulating, but do not exhibit charge ordering.  LaMnO$_3$ possesses A-type AFM order \cite{Wollen55} below the N\'{e}el temperature $T_N \approx 139$ K, in which ferromagnetic (FM) MnO$_2$ sheets are antiferromagnetically aligned with one another.  Conversely, CaMnO$_3$ possesses G-type AFM order \cite{Wollen55} below  $T_N \approx 123$ K, where each Mn atom is antiferromagnetically aligned with its nearest neighbor.  

Figure~\ref{fig:lamno1} shows the temperature dependence of the two muon frequencies observed in LaMnO$_3$ in zero applied field, together with their  amplitudes and fractional linewidths $1/(2\pi\nu_{\mu 1}T_{2i})$.  These two frequencies have the same temperature dependence, and  their signal amplitudes $(A_{\rm osc})_i$  are essentially the same and independent of temperature.  Note that the fractional linewidths below $T_N$ are very narrow ($\leq 2$ \%), as seen in the bottom frame of Fig.~\ref{fig:lamno1}. 

Figure~\ref{fig:lamno2} shows the temperature dependence of the spin-lattice relaxation rate $1/T_1$ and the relaxing amplitude $A_{\rm rlx}$ in LaMnO$_3$.  These data were obtained by fitting the relaxing part of $G_z(t)$ to Eq. (3).  The exponent $K$ was found to be consistent with a simple exponential relaxation function ($K = 1$) over the entire temperature region measured.   The sharp rise in $1/T_1$ at $T_N$ and fall just below T$_N$ indicates a very narrow range of critical  temperatures  for this compound. The magnitude of $1/T_1$ is temperature independent above $T_N$. Note also that $A_{\rm rlx}$ is equal to one above $T_N$, but falls within a few degrees to its polycrystalline average of about $1/3$ just below $T_N$, indicating that the transition to the AFM state is quite sharp in temperature.  Both the rapid fall in $A_{\rm rlx}$ and the sharp rise in $1/T_1$ occur very near the temperature where the observable muon frequencies approach zero.  The latter is shown in the bottom frame of Fig.~\ref{fig:lamno2}, where the two muon frequences have been normalized at $120$ K to show that they have the same temperature dependence.  Finally, the top frame of Fig.~\ref{fig:lamno2} shows that a rather modest field of $3$ kOe applied parallel to the muon spin polarization destroys the sharp critical behavior in $1/T_1$.

Figures~\ref{fig:camno1} and \ref{fig:camno2} show comparable data for CaMnO$_3$.  In this compound three muon frequencies are observed at low temperatures, but only the lowest frequency line $\nu_{\mu1}$ is clearly observable at temperatures above about $90$ K. The magnitude of $\nu_{\mu 1}$ approaches zero frequency near $T_N$, as seen in the middle frame of Fig.~\ref{fig:camno1} and the bottom frame of Fig.~\ref{fig:camno2}.  The relative linewidths $1/(2\pi \nu_{\mu i} T_{2i})$ are somewhat broader in CaMnO$_3$ than in LaMnO$_3$.  The dynamical signal in CaMnO$_3$ is shown in the top two frames of Fig.~\ref{fig:camno2}, and exhibits a behavior similar to that in LaMnO$_3$; e.g., a very narrow critical region, as indicated by sharp transitions in $1/T_1$ and $A_{\rm rlx}$, and a temperature-independent $1/T_1$ behavior above $T_N$. 

Summarizing our results on the two end compounds, LaMnO$_3$ and CaMnO$_3$ exhibit sharp transitions into their magnetically ordered states, and their dynamic relaxation functions $G_{\rm rlx}(t)$ can be described by a single exponentially decaying time dependence.   

\subsection{La$_{0.94}$Ca$_{0.06}$MnO$_3$: Insulating Canted Antiferromagnet}
As Ca is substituted for La one introduces chemical disorder into the lattice and holes into the Mn$^{3+}$ electronic structure, producing Mn$^{4+}$.  An additional consequence is that the local JT distortions are reduced as the holes become more abundant, because Mn$^{4+}$ is not a JT-active ion.  In La$_{0.94}$Ca$_{0.06}$MnO$_3$, for example, the AFM order found in LaMnO$_3$ begins to show traces of FM, although only through a canting of the Mn spin moments.\cite{Wollen55}  The system remains insulating, although its resistivity  drops in magnitude compared to the undoped LaMnO$_3$ (Fig.~\ref{fig:resist}).  

The $\mu$SR data for this system are now discussed.  First, no oscillating component was observed.  This is likely due to a complicated magnetic structure which washes out observation of a discreet muon frequency or frequencies.  The $\mu$SR data for the relaxing component $G_{\rm rlx}$ in La$_{0.94}$Ca$_{0.06}$MnO$_3$ are shown in Fig.~\ref{fig:laca6mn1}.   These data were obtained by fitting $G_{\rm rlx}$ with the stretched-exponential form in Eq.~(\ref{eq:strexp}).  One sees that the amplitude $A_{\rm rlx}$ drops sharply and the spin-lattice relaxation rate $1/T_1$ shows a peak at $T \approx 122$ K, indicating the onset of  magnetic order.  Note that  Ca doping has decreased the transition temperature compared to the undoped LaMnO$_3$. Comparing the zero-field relaxation rates in La$_{0.94}$Ca$_{0.06}$MnO$_3$ to those in LaMnO$_3$ one notes that the width of the peak in $1/T_1$ is larger in the Ca-doped compound.  This is so  despite the fact that the transition width, as measured by the reduction in the relaxing amplitude $A_{\rm rlx}$ near the ordering temperature, remains quite narrow.  Furthermore, the relaxation function in La$_{0.94}$Ca$_{0.06}$MnO$_3$ is no longer exponential, as it was in LaMnO$_3$.  This is shown in the middle frame of Fig.~\ref{fig:laca6mn1}, where the temperature dependence of the exponent $K$ in Eq. (3) is displayed.  $G_{\rm rlx}$ actually begins to deviate from its exponential form ($K = 1$) at a temperature slightly above the peak temperature in $1/T_1$.  The value of $K$ continues to decline as the temperature is lowered below the transition.  As shown here and below, this non-exponential behavior and broadening of the transition in $1/T_1$ is characteristic of the  systems containing both La and Ca.  Finally, as in both the undoped LaMnO$_3$ and CaMnO$_3$ compounds, the application of a small applied field tends to destroy the signature of the magnetic ordering in the spin-lattice relaxation rate (Fig.~\ref{fig:laca6mn1}, bottom frame).

\subsection{La$_{0.82}$Ca$_{0.18}$MnO$_3$: Ferromagnetic Insulator}
The data in La$_{0.82}$Ca$_{0.18}$MnO$_3$,  in which  the  level of doping is  sufficient to induce ferromagnetism, is now discussed.   As shown in Fig.~\ref{fig:resist}, the magnitude of the resistivity in this compound is lower than that of the La$_{0.94}$Ca$_{0.06}$MnO$_3$ material, indicating that the addition of doped holes is pushing the system closer to a conducting state.  The material remains insulating, however.  According to the phase diagram of Cheong and Hwang\cite{phasediagram} this material is a FM insulator with a transition to a CO state below about $60$ K.  Note that the resistivity for this compound shows a marked upturn at temperatures below about $80$ K (see Fig.~(\ref{fig:resist})).  

Figure~\ref{fig:laca18mn1} shows the temperature dependence of $A_{\rm rlx}$, $K$ and $1/T_1$ resulting from a fit to a stretched-exponential function for $G_{\rm rlx}$.  A sharp peak in the spin-lattice relaxation rate is evident at temperatures near $182$ K, and a broader peak is evident at lower temperatures near $110$ K.  The sharpness of the peak at $182$ K is characteristic of critical slowing down near a magnetic phase transition.  This transition to an ordered state is also reflected in the sharp reduction in the magnitude of $A_{\rm rlx}$ below the same temperature.  The peak temperature is consistent with the ferromagnetic transition temperature in the published phase diagram,\cite{phasediagram} indicating that  $T_C = 182$ K for this material.  As in the La$_{0.94}$Ca$_{0.06}$MnO$_3$ compound, the exponent $K$ decreases at temperatures below $T_C$. 

Figure~\ref{fig:laca18mn2} shows the temperature dependence of the measured muon frequency $\nu_\mu$, together with its  fractional linewidth $1/(2 \pi \nu_{\mu} T_2)$  from the oscillating portion of $G_{\rm rlx}$.  For comparison, the temperature dependence of $1/T_1$, shown in the bottom panel of the same figure, exhibits a peak where $\nu_{\mu}$ tends to zero.  Note that only a single frequency is observed,  unlike in LaMnO$_3$ or CaMnO$_3$, which display multiple muon frequencies.  This indicates that the Ca doping is sufficient to have created a rather simple (FM) magnetic structure, so that a discrete muon frequency is observable, unlike the case of La$_{0.94}$Ca$_{0.06}$MnO$_3$.  This observable frequency clearly corresponds to the buildup of the FM  magnetization below $T_C$, but observation of the FM order parameter is only visible in the temperature range $150 \leq T \leq 180$ K.  At the lowest temperatures where $\nu_\mu$ can be observed, $T \approx 0.82T_C$, the fractional spread in frequencies ($1/(2 \pi \nu_{\mu} T_2)$) is at least an order of magnitude larger than found in the undoped LaMnO$_3$.  This presumably reflects the local disorder in the material, which shows up as a spread in local magnetic fields, even though the transition temperature itself remains fairly sharp.  

Below $150$ K the ability to resolve a discrete muon frequency in La$_{0.82}$Ca$_{0.18}$MnO$_3$ vanishes, possibly due to an increasing linewidth or to a change in the magnetic state. At still lower temperatures, the muon relaxation rate $1/T_1$ shows a broad maximum, indicating a change in the Mn-ion spin dynamics. To observe this broad maximum  it was necessary to freeze the value of $K$ in the least-squares fitting procedure so that the form of the relaxation function would remain the same over the temperature range below about $170$ K.  

\subsection{La$_{0.67}$Ca$_{0.33}$MnO$_3$: Ferromagnetic Conductor}
For La$_{0.67}$Ca$_{0.33}$MnO$_3$ the Ca concentration is sufficient to produce a conducting state below the FM transition temperature $T_C \approx 270$ K (see Fig.~\ref{fig:resist}).  These data are from the same sample as used previously;\cite{Heffner96} however, the current data are of greater statistical precision than those previously reported and were taken at smaller temperature intervals near $T_C$.  Figures~\ref{fig:laca33mn1} and \ref{fig:laca33mn2} show the parameters obtained from fitting the relaxation function $G_z(t)$ using the stretched exponential form in Eq. (3) for $G_{\rm rlx}$.  Only a single muon frequency $\nu_\mu$ is found, the temperature dependence of which is shown in the middle frame of Fig.~\ref{fig:laca33mn1}.  The fractional inhomogeneous linewidth $1/(2\pi \nu_\mu T_2)$ is independent of temperature and yields about the same fractional width as in the insulating FM compound La$_{0.82}$Ca$_{0.18}$MnO$_3$.    The spin-lattice relaxation rate $1/T_1$ is plotted in  Fig.~\ref{fig:laca33mn2}, together with the exponent $K$ and the amplitude $A_{\rm rlx}$.  The latter shows that the FM transition is not as sharp in this material as in the previously discussed compounds.  Furthermore, the exponent $K$ falls dramatically below $T_C$, reaching values as low as about 0.2.   Finally, the temperature dependence of $1/T_1$ reveals a rather broad peak, with large relaxation rates extending over a temperature range of nearly $75$ K below $T_C$.  This persistence of a sizeable relaxation rate below the magnetic ordering temperature is more prominent at this Ca concentration than at any other presented here.  Finally, a small applied magnetic field of only $1 - 3$ kOe destroys the peak in $1/T_1$ near $T_C$.

\subsection{La$_{0.33}$Ca$_{0.67}$MnO$_3$: Charge Ordering and Antiferromagnetism}
When $0.50 \leq x \leq 0.88$ in La$_{1-x}$Ca$_{x}$MnO$_3$ the system undergoes a CO transition, followed by AF order at lower temperatures.\cite{phasediagram}  The highest CO transition temperature appears to occur for $x \approx 2/3$ where $T_{co} \approx 270$ K and $T_N \approx 125-150$ K.  These numbers are determined largely from dc susceptibility and resistivity measurements.\cite{phasediagram}  The $\mu$SR relaxation data between $5 \leq T \leq 250$ K for  La$_{0.33}$Ca$_{0.67}$MnO$_3$ are shown in Fig.~\ref{fig:laca67mn1}, again using a stretched exponential form for $G_{\rm rlx}$.  No $\mu$SR oscillations were observed in this material.  The temperature dependence of the spin-lattice relaxation rate $1/T_1$ shows two peaks, one near the N\'{e}el temperature $T_N \approx 150$ K and a lower, broader peak centered around $80$ K.  The temperature dependence of the relaxing asymmetry $A_{\rm rlx}$ shows a decrease from 100\% to about 33\% near 150K, which is the signature of magnetic order in a polycrystalline material.  This onset of magnetic order coincides with the peak in $1/T_1$, but the decrease in $A_{\rm rlx}$ occurs over a $20 - 30$ K temperature range, which is comparable to that in La$_{0.67}$Ca$_{0.33}$MnO$_3$, but considerably larger than in the lower-doped samples discussed above.   Also, similar to the observations in the previously discussed Ca-doped compounds, the exponent $K$ decreases from about $1$ above $T_N$  to $K \leq 0.4$ below $T_N$.

\subsection{Ferromagnetic Materials La$_{0.82}$Ca$_{0.18}$MnO$_3$ and La$_{0.67}$Ca$_{0.33}$MnO$_3$: A two-exponential analysis}
As mentioned above, the stretched-exponential relaxation function is characterized by only a few parameters and can therefore provide a useful parameterization of the data over a broad temperature range.  For example, a deviation from the expected\cite{Turov} and usually observed\cite{Gubbens94} simple exonential form ($K = 1$) is readily detectable.  However, careful examination of the data in Figs.~\ref{fig:laca33mn1} and \ref{fig:laca33mn2} raises some concerns about the particular applicability of this function in La$_{0.67}$Ca$_{0.33}$MnO$_3$.   First, the exponent $K$ reaches a very small value just below $T_C$, where $K \leq 0.2$.  Even in spin glasses such as {\em Ag}Mn, where a stretched exponential $\mu$SR relaxation function is observed, the exponent $1/3 \leq K \leq 1$.\cite{spinglass}  (In {\em Ag}Mn a distribution of energy barriers for  Mn spin rotation yields a broad distribution of correlation times $\tau$.)  Second, in La$_{0.67}$Ca$_{0.33}$MnO$_3$ the peak in $1/T_1$ actually occurs at a temperature {\em below} $T_C$  (Figs.~\ref{fig:laca33mn1} and \ref{fig:laca33mn2}), rather than being at $T_C$ as expected.  This effect is not seen in the other materials and  most likely occurs because, in the temperature region where $K$  is rapidly changing, the $1/T_1$ values at different temperatures are not derived from the same functional form.   These issues suggest that a more refined analysis may be useful.

Under ideal conditions determination of the precise shape of a decaying curve requires observation of the decay function over many decades with very high statistics.\cite{Istratov99}   Although  $\mu$SR data do not always meet these stringent criteria, we find that  the relaxation function $G_{\rm rlx}$ can  be better approximated by a sum of two exponentials (Eqn. 4), rather than a stretched exponential, when $K < 1$  and $1/T_1$ is reasonably large (usually $\geq 0.1 \mu s^{-1}$).     These conditions are realized for the two ferromagnetic materials La$_{0.82}$Ca$_{0.18}$MnO$_3$  and La$_{0.67}$Ca$_{0.33}$MnO$_3$, over a limited temperature range.  This is illustrated in Fig.~\ref{fig:laca33mn3}, which displays  $G_{\rm rlx}(t)$ in La$_{0.67}$Ca$_{0.33}$MnO$_3$.  The top frame shows least squares fits for a simple exponential ($K = 1$) and a stretched exponential ($K = 0.63$).  Clearly the stretched exponential function yields a better fit to the data.  The bottom frame in this figure shows a fit to the sum of two exponentials using Eqn. 4;  one can see at early times that the two-exponential fit is slightly better than that using the stretched exponential form.  

To more precisely illustrate the two-exponential nature of $G_{\rm rlx}$ in La$_{0.67}$Ca$_{0.33}$MnO$_3$ near $T_C$, in Fig.~\ref{fig:laca33mn4} we plot $-t/\ln(G_{\rm rlx})$ versus $t$ on a log-log plot, together with the calculated curves from the fits to the stretched- and two-exponential functions shown in Fig.~\ref{fig:laca33mn3}.   Plotted in this way, a purely exponential relaxation function yields a horizontal line (zero slope); as the exponent $K$ of the stretched-exponential is decreased from one, the calculated curve remains linear with slope $(1 - K)$. Clearly, the two-exponential model function yields a  better representation of the data. An analysis of the relaxation function in La$_{0.82}$Ca$_{0.18}$MnO$_3$ yields similar results. 

As stated above, fitting a relaxing curve to the sum of two or more exponentials is not rigorously unique.  Credible results are obtained for Ca concentrations $x = 0.18$ and $0.33$ when $K < 1$ and $1/T_1 \geq 0.1 \mu$s$^{-1}$, however, because the two rates $\lambda_f$ and $\lambda_s$ in Eq. 4 differ by more than a factor of ten.  This is illustrated in Fig.~\ref{fig:asymm}.  For La$_{0.92}$Ca$_{0.08}$MnO$_3$  and  La$_{0.33}$Ca$_{0.67}$MnO$_3$, although a stretched-exponential relaxation function fits the data reasonably well, the rates $1/T_1$ were too small to yield good, convergent two-exponential fits over a  temperature range wide enough to produce interpretable results.

Figs.~\ref{fig:laca18mnfs}  and \ref{fig:laca33mnfs} show the temperature dependence of the parameters $\lambda_f$ and  $\lambda_s$ obtained from fitting $G_{\rm rlx}$ in La$_{0.82}$Ca$_{0.18}$MnO$_3$ and La$_{0.67}$Ca$_{0.33}$MnO$_3$, respectively, using Eqs. 3 and 4.    The two-exponential fits could only be performed over a limited temperature range above and below $T_C$, where the relaxation rate $1/T_1$ is large enough, as explained above.  Sufficiently far above $T_C$, where $K \approx 1$, the data are well represented by a single exponential function. The open symbols correspond to the temperature region where two-exponential fits were feasible.  The closed symbols at higher temperatures (top frames Figures~\ref{fig:laca18mnfs}  and \ref{fig:laca33mnfs}) are from the single exponential fits.  In order that all of the plotted data for La$_{0.67}$Ca$_{0.33}$MnO$_3$ in the top frame of Fig.~\ref{fig:laca33mnfs} correspond to an exponential functional form, some data near $180$ K were omitted from this plot, because a two-exponential fit was not feasible and  the exponent $K$ in Eq. 3 was significantly less than $1$.  The temperature dependence of the parameters $A_f$ and $A_s$ are shown in Fig.~\ref{fig:lacaamps}. Here $A_f + A_s = A_{\rm rlx}$.

Analyzed in this way the data for La$_{0.82}$Ca$_{0.18}$MnO$_3$ and La$_{0.67}$Ca$_{0.33}$MnO$_3$ yield very interesting results. First we note that in both materials only $\lambda_f$ shows a peak at $T_C$, characteristic of the slowing down of spins near their magnetic ordering temperature.  The temperature dependence of $\lambda_s$ shows no such peak, displaying only temperature independent behavior below $T_C$.  Furthermore, the relaxation rates $\lambda_f$ and $\lambda_s$ differ by at least an order of magnitude, with $\lambda_s \approx 40 \lambda_f$ below $T_C$.  The ratio of amplitudes $A_f/A_s$ near and below $T_C$ is  $\approx 3.0$ in La$_{0.82}$Ca$_{0.18}$MnO$_3$ and $\approx 1.5$ in La$_{0.67}$Ca$_{0.33}$MnO$_3$.  Finally, as seen in Fig.~\ref{fig:lacaamps}, the amplitudes remain approximately independent of temperature below $T_C$ in  La$_{0.82}$Ca$_{0.18}$MnO$_3$, whereas $A_f$ increases and $A_s$ decreases below $T_C$ in La$_{0.67}$Ca$_{0.33}$MnO$_3$. 

\section{Analysis and Discussion of Results}
\subsection{Muon lattice sites}
The observation of two muon frequencies below $T_N$ in LaMnO$_3$  indicates that the muon occupies magnetically inequivalent sites in this material.  A similar situation was observed by  
Holzschuh {\em et al.\/}, \cite{Holz} who used $\mu$SR to study the muon's behavior in a series of single-crystal orthoferrite REFeO$_3$ (RE = rare earth) materials having the same crystal symmetry as the (La,A)MnO$_3$ samples used in our experiments.  Holzschuh {\em et al.\/} found that  the muon occupies three different crystalline sites at low temperatures in both ErFeO$_3$ and YFeO$_3$ but that, due to thermal activation, only one of these sites remains stable at temperatures above $100$ and $300$ K, respectively.  This most predominant and stable site is located about 1 \AA~ from the O(1) oxygen atoms (as defined below) in the REO$_2$ plane. 

As stated earlier the observed muon frequency $\nu_\mu$ is given by $(\gamma_\mu/2\pi)B$, where  the local magnetic field $B$  arises from the sum of a transferred hyperfine field $B_{hyp}$ and a local dipolar field $B_{dip}$.   In oxides, where the muon is bound about $1$ \AA~ from an oxygen atom, the hyperfine field is in part due to the formation of a muon-oxygen covalent bond.\cite{Boekema}  Holzschuh {\em et al.\/} found that their data in the orthoferrites were consistent with $B_{hyp} \ll B_{dip}$, however.  A similar finding has been reported from $\mu$SR experiments in the cuprates.\cite{cuprates}  It is therefore reasonable to assume that $B_{hyp}$ is also negligible in the manganites.  

Using the recently measured\cite{proffen} orthorhombic structure of LaMnO$_3$, where the Mn sit at ($0, 0.5, 0$) and the oxygen O(1) at ($0.073, 0.485, 0.25$) and O(2) at ($0.224, 0.304, 0.039$), we calculated the dipole fields in the unit cell to search for possible muon sites in a locus of about $1$ \AA~ from either the O(1) or O(2) sites.  Here the  atomic coordinates are given in fractions of the lattice spacings ($5.542$~\AA, $5.732$~\AA, $7.783$~\AA).  The calculated dipole fields were compared to the measured low-temperature frequencies in LaMnO$_3$, La$_{0.67}$Ca$_{0.33}$MnO$_3$ and CaMnO$_3$.  In LaMnO$_3$ there are two observed low-temperature frequencies, $\nu_{\mu1} =  84.90$ MHz and $\nu_{\mu2} = 128.8$ MHz, corresponding to local magnetic fields of $1.6$ kOe/$\mu_B$ and $2.4$ kOe/$\mu_B$, respectively.  We find two muon sites which are consistent with these data.  The first site [Mu(1), at ($0.06, 0.71, 0.25$)] is about $1$ \AA~ from O(1) and is quite close to the Holzschuh site found for the orthoferrites.  The second site [Mu(2), at ($0.38, 0.34, 0.0$)] is about $1$ \AA~ from O(2).  The calculated dipole fields in LaMnO$_3$ (Mn spin $S = 2$) for the known AFM structure  are $1.6$ and $2.4$ kOe/$\mu_B$ for Mu(1) and Mu(2), respectively, which are in excellent agreement with the measured values.  Furthermore, these same two sites reproduce the single measured low-temperature frequency in FM La$_{0.67}$Ca$_{0.33}$MnO$_3$, $\nu_{\mu} = 75$ MHz, or $1.5$ kOe/$\mu_B$; i.e., the calculated value of $B_{dip}$ is $1.5$ kOe/$\mu_B$ for {\em both} of these sites, using $S = 2(1-x) + 1.5x, x = 0.33$.  Finally, in CaMnO$_3$ three muon frequencies are observed: $\nu_{\mu1} =  14.3$ MHz, $\nu_{\mu2} = 41.2$ MHz and $\nu_{\mu3} = 86.5$ MHz, corresponding to $0.35$ kOe/$\mu_B$,  $1.01$ kOe/$\mu_B$, and $2.13$ kOe/$\mu_B$, respectively.  The calculated dipolar fields yield $B_{dip} = 0.27$ kOe/$\mu_B$ for Mu(2) and $B_{dip} = 2.3$ kOe/$\mu_B$ for Mu(1) using $S = 1.5$, in good agreement with $\nu_{\mu1}$ and $\nu_{\mu3}$, respectively.  Thus, these two muon sites are consistent with all of the measured low-temperature muon frequencies in LaMnO$_3$, La$_{0.67}$Ca$_{0.33}$MnO$_3$ and CaMnO$_3$ except the $\nu_{\mu2}$ frequency in CaMnO$_3$, which corresponds to a metastable position, as discussed below.  

A more definitive identification of the muon sites in these materials would require single crystals, in which not only the magnitude but also the direction of the local field could be determined.   Nevertheless, from the measurements reported here one may strongly infer that the positive muon occupies positions within about $1$ \AA~ of both the O(1) and O(2) oxygen atoms across the (La,Ca)MnO$_3$ series.  Because a small amount of local lattice dilation due to the electrostatic field of the muon can reduce the calculated dipole fields by about $10$\%, and because we neglect the possible effects of any small transferred hyperfine fields, we estimate that our position assignments are  uncertain to at least $0.2$ \AA.  Although knowledge of the muon site is very important for some types of experiments, we point out that the interpretation of our  primary results, the Mn-ion relaxation rates, does not depend on the exact location of the muon in the lattice.

It is important to know whether the muon is stationary in the lattice, however, because rapid muon motion can mimic the relaxation of the Mn spins themselves. The occurence of stable muon sites does not apply to CaMnO$_3$, for example.  Here three muon frequencies were observed, with only the lowest frequency $\nu_{\mu 1}$ persisting at temperatures as high as  $T_N$.  The amplitude corresponding to $\nu_{\mu 1}$  also increases monotonically with temperature, as seen in the top frame of Fig.~\ref{fig:camno1}.  This is likely due to a change in the muon site population with temperature, as found in the orthoferrites. 

A transfer of polarization between different muon sites occurs as a consequence of the thermal activation of the positive muon from a relatively shallow trapping site to a deeper one. \cite{Holz}  To affect the observed relative populations the muon hopping time $\tau_h$ must be sufficiently small; e.g.,   significant population change takes place at a temperature where $\tau_h \Delta\omega_\mu \leq 1$,  with $\Delta\omega_\mu$ the frequency difference between the two sites.\cite{Holz}  While we do not attempt a detailed analysis of the muon motion in CaMnO$_3$, we estimate $\tau_h \cong 0.003~ \mu$s at $T \cong 80$ K in CaMnO$_3$, assuming $\Delta\omega_\mu \cong 2\pi \times 50$ MHz, corresponding to hopping between sites with frequencies $\nu_{\mu3}$ and $\nu_{\mu1}$, for example. We note, however, that the $1/T_1$ linewidth is independent of temperature above $T_N$ up to $275$ K.  This indicates that any postulated muon hopping rate out of the most stable site in CaMnO$_3$ (corresponding to $\nu_{\mu1}$) is small compared to the inverse of the Mn spin correlation time $\tau^{-1}$.  An order of magnitude estimate for $\tau^{-1}$ at high temperatures in CaMnO$_3$ yields $10^{11} - 10^{12}$ s$^{-1}$. Thus, we deduce that $3 \times 10^8 \leq \tau_h^{-1} \ll 10^{11} - 10^{12}$ s$^{-1}$ in CaMnO$_3$ for $T \geq T_N$.

Having obtained direct evidence for unstable muon sites in CaMnO$_3$, one must consider the possibility of muon motion in the other compounds, particularly at the highest measured temperatures.  Significant muon hopping is ruled out below $170$ K in LaMnO$_3$, for the following reasons.   Unlike in the orthoferrites or in CaMnO$_3$,  the two signal amplitudes which we measure in LaMnO$_3$ remain temperature independent below $T_N$ (top frame, Fig.~\ref{fig:lamno1}).  Thus these data, and the constancy of the spin-lattice relaxation rate above $T_N$,  indicate that the muon shows no sign of occupying a metastable site below at least $170$ K in LaMnO$_3$.   One might inquire whether the falling of the rate $\lambda_s$ in La$_{0.67}$Ca$_{0.33}$MnO$_3$ above about $267$ K (Fig.~\ref{fig:laca33mnfs}) could be caused by muon motion.  This would require a muon hopping rate $\tau_h^{-1} \gg \tau_s^{-1}$, the inverse correlation time of the \lq slow\rq~ Mn spins.  We  estimate $\tau_s^{-1} \approx 2 \times 10^{11}$ s$^{-1}$, where $\tau_s^{-1}$ is derived from the measured $\lambda_s \approx 2\mu$s$^{-1}$ and the relation $\lambda_s = 2\omega_\mu^2\tau_s$, using the low-temperature muon frequency $\omega_\mu = 2\pi \times 75$~MHz.  This yields a value for $\tau_h^{-1}$ in  La$_{0.67}$Ca$_{0.33}$MnO$_3$ which exceeds the upper limit found above for the most stable site in CaMnO$_3$ at comparable temperatures.  (This site, Mu(2), is also  found in La$_{0.67}$Ca$_{0.33}$MnO$_3$.) The disorder induced by doping in (La,Ca)MnO$_3$ should help to stabilize the muon position, and it therefore seems unlikely that muon motion is a significant factor in our measurements.  

\subsection{Muon precession frequencies}
The temperature dependence of the $\nu_\mu$ reflects the growth of the local magnetization below the ordering temperature.  We have performed an analysis of the temperature dependence of the muon frequency below the ordering temperatures in those materials where the muon frequency is observable.  The temperature dependence of $\nu_\mu (T)$ can be parameterized as follows:  
\begin{equation}
\nu_\mu(T) = \nu_{\mu 0} (1 - T/T_M)^\beta
\label{eq:mag}
\end{equation}
Fits were carried out in a limited temperature range near the ordering temperature $T_M$ ($0.80 T_M \lesssim T \lesssim T_M$), reflecting the limited range of validity for Eq.~(\ref{eq:mag}). (Strictly speaking Eq.~(\ref{eq:mag}) is valid only in the asymptotic critical region of a second order magnetic phase transition.  Often this expression works well outside of this region, however.)  The fits are shown as solid lines in the lower frames of Figures~\ref{fig:lamno2} and \ref{fig:camno2} for LaMnO$_3$ and CaMnO$_3$, respectively.  In LaMnO$_3$ the two frequencies were normalized at $T = 120$ K; one sees from the plot in Fig.~\ref{fig:lamno2} that the two signals possess the same temperature dependence.  We obtained $\beta = 0.340 \pm 0.008$ for LaMnO$_3$ and $\beta = 0.53 \pm 0.03$ for CaMnO$_3$; in the latter case only the more stable frequency $\nu_{\mu 1}$ was included in the fit. The exponent for LaMnO$_3$ is consistent with measured critical exponents in 3D Heisenberg antiferromagnets,\cite{critical} while the exponent for CaMnO$_3$ yields the mean field value of $1/2$. The presence of muon diffusion (i.e., change in muon site population) in CaMnO$_3$ may render this value unreliable, however. 

The fit for the La$_{0.82}$Ca$_{0.18}$MnO$_3$ material, carried out over the observable temperature range ($0.8 \lesssim T/T_C \lesssim 1.0$), is shown in the middle frame of Fig.~\ref{fig:laca18mn2}.  This yielded $T_C = 182.2 \pm 1.8$ K and $\beta = 0.49 \pm 0.09$.  A fit to the La$_{0.67}$Ca$_{0.33}$MnO$_3$ data between $T = 220 - 267$ K yielded $T_C = 268 \pm 0.4$ K and $\beta = 0.233 \pm 0.013$.\cite{note1}    

In each of these cases the fitted frequency curves give a magnetic ordering temperature which is consistent with the peak in the spin-lattice relaxation rates arising from critical slowing down of the Mn spins.  Note that for La$_{0.67}$Ca$_{0.33}$MnO$_3$ this is only true for the two-exponential fit, not for the stretched exponential fit.  This fact is further indication that the two-exponential fit is a better representation of the data near $T_C$ for this system.  

It is not clear how to interpret the values of $\beta$ obtained from the FM materials $x = 0.18$ and $0.33$, for the following reasons.  The insulating $x = 0.18$ material lies close to the boundary where canted AFM is observed.  Furthermore, the growth of the FM order parameter in this compound may be influenced by the change in spin dynamics found below $150$ K, near where the resistivity shows its anomalous increase.  Either of these facts makes comparison of  $\beta$ with theories of conventional phase transitions dubious.  Finally, as  discussed below, the transition in the $x = 0.33$ conducting material may proceed via the percolation of FM clusters arising from the presence of magnetoelastic polarons, rather than via a more conventional FM second-order phase transition.  At present there is, therefore,  no theory to which these exponents can be compared. 
 
\subsection{Spin-lattice relaxation}
We first discuss the zero-field spin-lattice relaxation rates in LaMnO$_3$ and CaMnO$_3$, shown in the top frames of Figs.~\ref{fig:lamno2} and \ref{fig:camno2}, respectively.   In both LaMnO$_3$ and CaMnO$_3$ $G_{\rm rlx}$ is well described by a single exponential $A_{\rm rlx} \exp(-t/T_1)$.    The peak in the spin-lattice relaxation rate $1/T_1$ reflects critical slowing down near $T_N$, and  in each material the temperature region of the critical dynamics is relatively narrow.   Furthermore, the zero-field transition in both materials is quite sharp, as seen by the rapid decrease in the relaxing amplitude $A_{\rm rlx}$ below $T_N$, shown in the middle frames of Figs.~\ref{fig:lamno2} and \ref{fig:camno2}. Note that even the relatively modest ($\mu_B H \ll k_B T_N$) field $H = 3$ kOe applied parallel to the muon spin  strongly affects the critical dynamics. 

Both LaMnO$_3$ and CaMnO$_3$ display temperature independent spin-lattice relaxation over an extended temperature range above $T_N$.  This is characteristic of systems fluctuating at their exchange frequency $\omega_e$.  In this instance one has \cite{Turov}
\begin{equation}
1/T_1 = \sqrt{2 \pi} S (S + 1) \Delta^2 /(3 \hbar^2 \omega_e),
\label{eq:exchange}
\end{equation}
where $S$ is the Mn atomic spin ($S = 2$ for LaMnO$_3$ and $S = 3/2$ for CaMnO$_3$) and $\Delta$ is the hyperfine coupling constant between the muon spin and the spin $S$. The exchange frequency is related to the exchange integral $J$ via the expression $\hbar \omega_e = J \sqrt{2 z S (S + 1)/3}$, where $z$ is the number of nearest neighbor Mn spins to a given Mn spin. 

We now use Eq. (\ref{eq:exchange}) to estimate the exchange frequency for LaMnO$_3$ and CaMnO$_3$.  To do so one may obtain $\Delta$ from the value of the low temperature muon frequency ($2 \pi \nu_\mu = \omega_\mu$) or frequencies.  If there is a single frequency then $\Delta = \hbar \omega_{\mu}/S$.  When there is more than one muon frequency the relaxation function takes the form $G_{\rm rlx} = \Sigma_i (A_{\rm rlx})_i\exp(-t/T_{1i})$, where the index $i$ runs from $1$ to $3$ in CaMnO$_3$, for example.  In principle, this implies that one should observe a multi-component or stretched-exponential relaxation function, which is in contradiction to the  single-exponential function observed in both LaMnO$_3$ and CaMnO$_3$.  However, if the individual relaxation times $T_{1i}$ are all long compared to the characteristic time of the experiment (of the order of the muon lifetime), then 
\begin{equation}
G_{\rm rlx} = \Sigma_i (A_{\rm rlx})_i\exp(-t/T_{1i}) \approx A_{\rm rlx} \exp(-t/\overline{T_1}), 
\label{eq:T1i}
\end{equation}
where $1/\overline{T_1} = \Sigma_i (A_{\rm rlx})_i T_{1i}^{-1} / A_{\rm rlx}$.  In this limit a single exponential relaxation function is recovered, with a relaxation rate which is an appropriately weighted sum of rates $1/T_{1i}$.  The observed relaxation times $T_1$ in both LaMnO$_3$ and CaMnO$_3$ are relatively long  ($\approx 10 - 20  \mu$s), and thus the individual relaxation times $T_{1i}$ are also presumably long. This then explains the observation of an exponential relaxation function in each case.  

In the motional narrowing limit,\cite{Turov} appropriate for relaxation above $T_N$, $1/T_{1i} = 2 \omega_{\mu i}^2 \tau_i$, so that $\Delta^2 \approx \Sigma_i (A_{\rm rlx})_i (\hbar \omega_{\mu i}/S)^2/A_{\rm rlx}$.  Here  $\tau$ is the Mn-ion correlation time.    In LaMnO$_3$ the observed low temperature frequencies are $\omega_{\mu 1} = 2 \pi \times 84.9$ MHz and $\omega_{\mu 2} = 2 \pi \times 128.8$ MHz.  These values yield an exchange frequency $\omega_e = 6.33$ THz, corresponding to an exchange integral $J = 0.85$ meV.  Using the mean field expression\cite{Turov} $T_N = JzS(S + 1)/3$  one obtains $T_N = 119$ K, in good agreement with the measured Ne\`el temperature of $139$ K.  The situation is not so sanguine in CaMnO$_3$, however. Here the low temperature frequencies are $\omega_{\mu 1} = 2 \pi \times 14.3$ MHz, $\omega_{\mu 2} = 2 \pi \times 41.2$ MHz and $\omega_{\mu 3} = 2 \pi \times 86.5$ MHz, yielding $\omega_e = 2.8$ THz, $J = 0.47$ meV and $T_N = 42$~K, lower by a factor of about 3 than the measured $T_N$.  However, if one uses {\em only} the lowest frequency $\nu_{\mu 1}$ for CaMnO$_3$ (corresponding to the \lq stable\rq~ muon site), one obtains $\omega_e = 0.10$ THz, $J = 0.017$ meV and $T_N = 1.5$K, an unacceptably low value even given the crudeness of the approximations used. This seems to be an indication that in CaMnO$_3$ muon motion makes determination of the effective coupling constant $\Delta$ ambiguous.  Alternatively, the mean field approximations used to estimate $T_N$ may not be valid in CaMnO$_3$, despite the finding of a mean-field exponent for the temperature dependence of the lowest $\mu$SR frequency $\nu_{\mu1}$.

As the amount of Ca doping increases from zero in La$_{1-x}$Ca$_x$MnO$_3$ the width of the peak in $1/T_1$ near the magnetic transition grows larger.  This is seen already for $x = 0.06$  (Fig.~\ref{fig:laca6mn1}), where the width in $1/T_1$ extends over a considerably larger temperature region than does the fall in the relaxing amplitude $A_{\rm rlx}$.  Recall that a reduction in $A_{\rm rlx}$ signals the onset of magnetic order, even though no muon frequencies were observed in this material.  The changing value of the exponent $K$ also extends over a wide temperature regime around $T_N$, indicating that the spin dynamics are becoming inhomogeneous near $T_N$. This trend, a general decline in the value of $K$ at the magnetic ordering temperature,  is seen in all of the Ca-doped samples except the end compound CaMnO$_3$, in which simple exponential relaxation is observed at all temperatures.  Furthermore, the $1/T_1$ relaxation rates in the $x = 0.06$ and $x = 0.33$ materials tend to remain relatively large at temperatures significantly below their magnetic ordering temperatures, indicating that there is an anomalously slow (large $\tau$) relaxation process which persists  in the ordered  state.  Values of $K \ll 1$, combined with the real-space nature of the muon probe, suggest that  the relaxation rates are therefore spatially distributed, as discussed in more detail below.  

The FM  $x = 0.18$  compound and the AFM $x = 0.67$ compound each exhibit an uncharacteristic increase in their spin-lattice relaxation rates well below their ordering temperatures.  This change is marked by a broad maximum near $T \approx 110$~K for $x = 0.18$ ($T_C = 182$ K) and $T \approx 85$~K for $x = 0.67$ ($T_N \simeq 150$ K).  It is tempting to attribute the loss of the oscillating component and the gentle rise in the relaxation rates near $110$ K in La$_{0.82}$Ca$_{0.18}$MnO$_3$ to the influence of  \lq charge ordering\rq~(CO) reported by others\cite{phasediagram} at $T \approx 70$ K. It is known, for example, that the  dynamical spin correlations in (Bi,Ca)MnO$_3$ change from ferromagnetic above the CO temperature $T_{\rm co}$ to antiferromagnetic below $T_{\rm co}$.\cite{Bao}  We note, however, that CO observed in La$_{0.33}$Ca$_{0.67}$MnO$_3$  occurs at temperatures {\em above} $T_N$, as opposed to {\em below} $T_C$ in La$_{0.82}$Ca$_{0.18}$MnO$_3$.   Because $\mu$SR cannot determine the type of magnetic correlations present in these materials  (in general, this requires a momentum-dependent probe), we cannot definitively explain the change in spin dynamics below the ordering temperatures of the $x = 0.18$ and $x = 0.67$ materials.  We do note the possible correlation with CO in both, however.  This could be the subject of further investigation by neutron scattering, for example.

The two FM materials under study, La$_{0.82}$Ca$_{0.18}$MnO$_3$ and La$_{0.67}$Ca$_{0.33}$MnO$_3$, exhibit a bi-modal distribution of fluctuation rates in perhaps the most interesting part of the phase diagram. The two-exponential fits to $G_{\rm rlx}$ [Eq.~(\ref{eq:twoexp})] yielded relaxation rates and relative amplitudes $\lambda_f$, $\lambda_s$, $A_f$ and $A_s$.   We have labeled these quantities \lq f\rq~ and \lq s\rq~ to indicate \lq fast\rq~ and \lq slow\rq~ Mn-ion fluctuation rates $\tau^{-1}$ because, in general, the local muon relaxation rate $\lambda \propto \gamma^2_{\mu} \sum_q |\delta B(q)|^2 \tau(q)$, as mentioned in Section IA.  Here  $|\delta B(q)|$ is the amplitude of the local fluctuating field and $\tau(q)$ is the Mn-ion correlation time.  The relaxation rates  $\lambda_f$ and $\lambda_s$ display very different magnitudes (Figs.~\ref{fig:laca18mnfs} - \ref{fig:laca33mnfs}):  in each material $\lambda_s \approx 10-40 \times \lambda_f$. Using the relation $\lambda = 2\omega_{\mu}\tau$, one may derive Mn correlation times for La$_{0.67}$Ca$_{0.33}$MnO$_3$ at $T = 250$ K, for example.  One obtains $\tau_f \simeq 10^{-13}$ s and $\tau_s = 5 \times 10^{-12}$ s, assuming that $\omega_\mu$ at the lowest temperatures is representative of the $\mu^+$-Mn coupling for both the slow and fast components.  It is noteworthy that only $\lambda_f$ shows evidence of \lq critical slowing down\rq~ at $T_C$.  In the FM conductor La$_{0.67}$Ca$_{0.33}$MnO$_3$, $A_f$ increases below $T_C$ (and $A_s$ decreases), while in the insulating material La$_{0.82}$Ca$_{0.18}$MnO$_3$ the fast and slow amplitudes are relatively temperature independent below $T_C$ (Fig.~\ref{fig:lacaamps}).  At $T_C$, $A_f$ is also a somewhat larger in La$_{0.82}$Ca$_{0.18}$MnO$_3$ than in La$_{0.67}$Ca$_{0.33}$MnO$_3$.

This analysis is consistent with two different spatially distinguishable regions in the sample, characterized in our measurements by very different relaxation rates and temperature-dependent volumes.  More specifically, although the $\lambda_f$ component may exist throughout the sample volume, the $\lambda_s$ component can only exist in spatially separated regions; otherwise only the slow Mn fluctuation rate (which creates the largest $\mu$SR rate and corresponds to $\lambda_s$) would be observed.  The muon samples these separate regions locally through its relatively short-ranged dipolar coupling to the Mn spins (matrix element $\sim r^{-6}$).  It is  important to note that neither $\mu$SR relaxation component corresponds to pure LaMnO$_3$ or CaMnO$_3$, whose  relaxation rates are independent of temperature above 140 K with  magnitudes $\approx 0.10 - 0.05~ \mu s^{-1}$, as discussed earlier.  This argues against large-scale separation of Mn$^{3+}$ and Mn$^{4+}$ ions.  Furthermore, we note that because of the large spin wave stiffness constant observed\cite{lynn} near $T_C$ in these FM materials the muon spin is not significantly relaxed by spin waves, as pointed out\cite{Heffner96} previously.  Thus, these FM materials possess at least three types of fluctuations, the FM spin waves observed in neutron scattering, and the excitations associated with the $\lambda_f$ and $\lambda_s$ relaxation rates reported here.

We postulate that the two relaxation components observed here constitute the spin signature of magnetoelastic polarons as the system becomes FM.  As discussed in the introduction to this paper, these polarons are thought to be formed from FM polarized Mn spins which surround a region of local lattice distortion, and presumably originate from the combined DE and JT interactions. These polaronic effects, together with the random replacement of La atoms by Ca, produce considerable disorder.  Our measured $\lambda_f$ component shows a peak in the relaxation rate at $T_C$, and we associate this relaxation with {\em overdamped} spin waves, characteristic of a disordered FM.  That is, the damping of the spin waves in this scenario would be caused by a loss of translational symmetry arising from inherent disorder.  

As mentioned in the introduction, a broad peak in the neutron scattering intensity centered around zero energy transfer has been observed in FM La$_{0.67}$Ca$_{0.33}$MnO$_3$.\cite{lynn}  This peak coexists with Mn spin waves near $T_C$, and persists to temperatures as low as $T/T_C = 0.7$, consistent with the finite $\mu$SR rate we measure in La$_{0.67}$Ca$_{0.33}$MnO$_3$ below $T_C$.   The neutron linewidth $\Gamma$ obeys a diffusive relaxation law, $\Gamma = Dq^2$, where $D \approx 30$ mev \AA$^2$.  The length scale for this diffusion was found to be $12$ \AA~ from small angle scattering.\cite{lynn}  We can compare this linewidth $\Gamma$ with $\lambda_f$ using the relation\cite{Turov} 
\begin{equation}
\lambda_f = 2{\omega_\mu}^2 \int \langle {\bf S}(q,t) \cdot {\bf S}(-q,0) \rangle \exp(i\omega t) dt. 
\label{eq:lambdaf}
\end{equation}
Here $\langle {\bf S}(q,t) \cdot {\bf S}(-q,0) \rangle$ is the transverse spin-spin correlation function given by 
\begin{equation}
\langle {\bf S}(q,t) \cdot {\bf S}(-q,0) \rangle = k_B T  \chi(q)  \exp(-Dq^2 t)/(g \mu_b)^2, 
\label{eq:correl}
\end{equation}
where $\chi(q)$ is the static $q$-dependent susceptibility, $k_B$ is Boltzman's constant and $\mu_B$ is the Bohr magneton.  We take an Ornstein-Zernike form for $\chi(q)$, given by\cite{lovesey}
\begin{equation}
\chi(q) = (g \mu_b)^2 S(S + 1)/(3k_B T (q^2 + \xi^{-2})r_1^2), 
\label{eq:chi}
\end{equation}
where $r_1^2 = a^2/6$ for a cubic lattice with lattice constant $a$, and $\xi$ is the coherence length for the diffusive spin correlations.  Evaluating these expressions  for $\xi = 12$ \AA~ yields $\lambda_f = 0.18~ \mu$s$^{-1}$, in excellent agreement with our $\mu$SR measurements of $\lambda_f$ at $T = T_C$ (see Fig. \ref{fig:laca33mnfs}). Thus, we associate our observed $\lambda_f$ with the diffusive relaxation mode observed in quasi-elastic neutron scattering, and attribute the relaxation to {\em overdamped} spin waves. 
  
The existence of overdamped FM spin waves is naturally  associated with a relatively high level of charge mobility because the FM in these CMR compounds is induced by the hopping of holes via the DE interaction.  By contrast, the $\lambda_s$ component corresponds to a rather long correlation time of $\approx 10^{-11}$ s.  This order-of-magnitude is  probably inconsistent with electronic relaxation of single Mn$^{3+}$ or Mn$^{4+}$ spins.  Other possible mechanisms might include phonon induced spin relaxation (via magnetoelastic coupling) or  the slow overturning of spins associated with local polaron hopping in regions of the sample where spin and charge motion are frustrated by more extreme local lattice distortions, that is, regions which are relatively {\em insulating}.  

The  possiblity that the $\lambda_s$ component is associated with relatively insulating regions of the sample, while the $\lambda_f$ component is associated with relatively conducting regions,  is qualitatively consistent with the temperature dependence of the volume fractions $A_s$ and $A_f$ in the two FM materials.   In the FM conductor La$_{0.67}$Ca$_{0.33}$MnO$_3$ the magnitude of $A_f$ increases (and $A_s$ decreases) as the temperature is lowered below $T_C$ and the material becomes more conducting.  By contrast, in the insulating FM La$_{0.82}$Ca$_{0.18}$MnO$_3$ the volume fractions $A_f$ and $A_s$ are independent of temperature.  This suggests that even significantly below $T_C$ in  conducting La$_{0.67}$Ca$_{0.33}$MnO$_3$ there exist relatively non-conducting regions which are the remnants of the inhomogeneity produced by isolated polarons existing above $T_C$ in these materials. These less conducting regions could be caused by an uneven distribution of Ca atoms, or by more extreme local lattice distortions which inhibit charge motion.  As the temperature is lowered below $T_C$ in La$_{0.67}$Ca$_{0.33}$MnO$_3$ these less conducting regions are gradually absorbed into the larger conducting matrix of the material.  This apparently does not occur in the insulating FM La$_{0.82}$Ca$_{0.18}$MnO$_3$, however, where the volume fractions are frozen below $T_C$.  

Such a postulated scenario is consistent with our measurements, and with charge transport and structural measurements, as follows.   In the two-fluid model of the resistivity\cite{Jaime} growth of the conducting charge fraction in La$_{0.67}$Ca$_{0.33}$MnO$_3$ occurs rapidly as the magnetization grows, reaching essentially 100 \% just below $T_C$.  This does  not contradict the above picture from the $\mu$SR data (in which slowly fluctuating spins in relatively less metallic regions persist far below $T_C$) because when a conducting path is reached (at $T_C$) the resistivity is shorted out, even though a considerable volume fraction may still be relatively non-conducting. Thus the two-fluid model of the resistivity  is similar to the  picture presented here, with this one exception.   The persistence of an inhomogeneous ground state below $T_C$ is reinforced by local probes of the lattice structure,\cite{pdf,xafs} which are consistent with the {\em gradual} loss of structural inhomogeneity below $T_C$, similar to that reflected in the spin-lattice relaxation rates reported here in the conducting FM La$_{0.67}$Ca$_{0.33}$MnO$_3$.

\section{Summary}
In this paper we  present $\mu$SR data for a series of (La,Ca)MnO$_3$ compounds,  spanning the range from the AFMs LaMnO$_3$ and CaMnO$_3$, to the  FMs  La$_{0.82}$Ca$_{0.18}$MnO$_3$ and La$_{0.67}$Ca$_{0.33}$MnO$_3$, to the charge-ordered material La$_{0.33}$Ca$_{0.67}$MnO$_3$. The $\mu$SR technique allows one to obtain the temperature dependence of both the local static magnetic order parameters and the Mn spin-lattice-relaxation rates $\tau^{-1}$, where generally $10^{-4} \leq \tau^{-1} < 10^{-13}$ s$^{-1}$.  These measurements, therefore,  provide a necessary complement to other spin probes, such as neutron scattering and nuclear magnetic resonance.  

From the magnitude of the measured local magnetic fields we have identified two plausible stable muon sites near the O(1) and O(2) atoms in these materials.  Except for CaMnO$_3$, where evidence for metastable muon sites is found, we conclude that rapid muon diffusion is not a factor in our measurements of the Mn spin dynamics.  We find that the spin dynamics near the AFM and FM magnetic ordering temperatures of these compounds is quite sensitive to small applied magnetic fields, however.  This may be a  consequence of the FM double-exchange interaction, which is enhanced by the application of a uniform field.  From the magnitude of the zero-field $\mu$SR rate above $T_N$ in the AFM insulating parent compound LaMnO$_3$ we derive an exchange integral $J = 0.83$ meV, which is consistent with the mean field expression for $T_N$.  Our principal finding is that the addition of Ca to LaMnO$_3$ (or La to CaMnO$_3$) results in an inhomogeneous spectrum of zero-field $\mu$SR rates below the magnetic ordering temperatures in these materials.  In the FM compounds La$_{0.82}$Ca$_{0.18}$MnO$_3$ and La$_{0.67}$Ca$_{0.33}$MnO$_3$ we are able to resolve this inhomogeneity into a bi-modal distribution of \lq fast\rq~ and \lq slow\rq~ $\mu$SR rates, corresponding to Mn spin fluctuation rates which differ by about a factor of at least $30$.  Because $\mu$SR  is a local probe this implies that the \lq slow\rq~ Mn fluctuations must exist in distinct spatial regions of the sample.  A physical picture is hypothesized for these FM phases in which the fast Mn rates  are due to overdamped spin waves characteristic of a disordered FM, and the  slower Mn relaxation rates  derive from distinct, relatively insulating regions in the sample.  These results present direct evidence for microscopic inhomogeneity in the spin channel of the FM manganites, suggesting that a proper description of these materials probably includes more than just the simple JT and DE interactions.

This work was supported in part by the U.S. National Science Foundation, Grants DMR-9731361 (Riverside), and DMR-9510454 (Columbia),  by  the Japanese agency NEDO (Columbia), and the Netherlands agencies FOM and NWO (Leiden). The research was carried out in part under the auspices of the U.S. DOE (Los Alamos). This work was supported by the Director for Energy Research, Office of Basic Energy Sciences.

\break
\begin{figure}
\caption{Temperature dependence of the $\mu$SR oscillating amplitude (top), precession frequencies (middle) and fractional linewidth (bottom) in LaMnO$_3$ from fitting to Eq.~(\ref{eq:Gosc}).}
\label{fig:lamno1}
\end{figure}

\begin{figure}
\caption{Temperature dependence of the $\mu$SR spin-lattice relaxation rate (top), relaxing amplitude (middle) and muon precession frequencies $\nu_{\mu 1}$ and $\nu_{\mu 2}$ (bottom)in LaMnO$_3$. $\nu_{\mu 2}$ has been normalized to $\nu_{\mu 1}$ at $120$ K.    The solid curve in the bottom frame is from a fit of the precession frequencies to Eq.~(\ref{eq:mag}).}
\label{fig:lamno2}
\end{figure}

\begin{figure}
\caption{Temperature dependence of the $\mu$SR oscillating amplitude (top), precession frequencies (middle) and fractional linewidth (bottom) in CaMnO$_3$ from fitting to Eq.~(\ref{eq:Gosc}).}
\label{fig:camno1}
\end{figure}

\begin{figure}
\caption{Temperature dependence of the $\mu$SR spin-lattice relaxation rate (top), relaxing amplitude (middle) and muon precession frequency $\nu_{\mu 1}$ (bottom) in CaMnO$_3$.    The solid curve in the bottom frame is from a fit of  $\nu_{\mu 1}$ to Eq.~(\ref{eq:mag}).}
\label{fig:camno2}
\end{figure}

\begin{figure}
\caption{Temperature dependence of the resistivity for La$_{1-x}$Ca$_x$MnO$_3$, $x = 0.0, 0.06, 0.18$ and $0.33$.  Comparable data for the $x = 0.67$ sample are given in S-W. Cheong and C. H. Chen, {\em Colossal Magnetoresistance and Related Properties}, ed. B. Raveau and C. N. R. Rao (World Scientific), to be published.}
\label{fig:resist}
\end{figure}

\begin{figure}
\caption{Temperature dependence of the $\mu$SR relaxing amplitude (top), exponent (middle) and spin-lattice relaxation rate (bottom) in La$_{0.94}$Ca$_{0.06}$MnO$_3$ obtained from a fit to Eq.~(\ref{eq:strexp}).  The filled symbols are for zero applied field, and the open circles for an applied field of $2.5$ kOe.}
\label{fig:laca6mn1}
\end{figure}

\begin{figure}
\caption{Temperature dependence of the $\mu$SR relaxing amplitude (top), exponent (middle) and spin-lattice relaxation rate (bottom)in La$_{0.82}$Ca$_{0.18}$MnO$_3$ obtained from a fit to Eq.~(\ref{eq:strexp}).}
\label{fig:laca18mn1}
\end{figure}

\begin{figure}
\caption{Temperature dependence of the $\mu$SR fractional linewidth (top), precession frequency (middle) and spin-lattice relaxation rate (bottom)in La$_{0.82}$Ca$_{0.18}$MnO$_3$. The data in the top and middle frames are from a fit to Eq.~(\ref{eq:Gosc}). The solid line in the middle frame is from a fit of $\nu_\mu$ to Eq.~(\ref{eq:mag}).} 
\label{fig:laca18mn2}
\end{figure}

\begin{figure}
\caption{Temperature dependence of the $\mu$SR fractional linewidth (top), precession frequency (middle) and spin-lattice relaxation rate (bottom) in La$_{0.67}$Ca$_{0.33}$MnO$_3$ obtained  from a fit to Eq.~(\ref{eq:strexp}). The inset  in the middle frame shows a fit of $\nu_\mu$ to Eq.~(\ref{eq:mag}).} 
\label{fig:laca33mn1}
\end{figure}

\begin{figure}
\caption{Temperature dependence of the $\mu$SR relaxing amplitude (top), spin-lattice relaxation rate (middle) and exponent (bottom) for zero applied field (filled symbols) in La$_{0.67}$Ca$_{0.33}$MnO$_3$ obtained from  fits to Eq.~(\ref{eq:strexp}). The open symbols in the middle frame show data in applied magnetic fields of $1$ kOe and $3$ kOe.}
\label{fig:laca33mn2}
\end{figure}

\begin{figure}
\caption{Temperature dependence of the $\mu$SR relaxing amplitude (top), spin-lattice relaxation rate (middle) and exponent (bottom) for zero applied field  in La$_{0.33}$Ca$_{0.67}$MnO$_3$ obtained from  fits to Eq.~(\ref{eq:strexp}).}
\label{fig:laca67mn1}
\end{figure}

\begin{figure} 
\caption{(a) $\mu$SR relaxation function $G_{\rm rlx}(t)$ at $T = 270$ K.  The  curves show best fits using stretched exponential from Eq.~(\ref{eq:strexp}) [$1/T_1 = 0.219(7)~ \mu s^{-1},~K = 0.63(3)$] and exponential ($K = 1$) functions. (b) A two-exponential least squares fit using Eq.~(\ref{eq:twoexp}), with $A_f = 0.60(5)$, $A_s = 0.40(5)$, $\lambda_f = 0.106(5)~ \mu s^{-1}$ and $\lambda_s = 0.93(11)~ \mu s^{-1}$.}
\label{fig:laca33mn3}
\end{figure}

\begin{figure} 
\caption{The measured $\mu$SR relaxation function $G_{\rm rlx}(t)$ at $T = 270$ K (open squares), normalized such that $G_{\rm rlx}(0) = 1$.  The   lines show calculated stretched-exponential (lower dotted line) and two-exponential (upper dashed line) functions using the fit parameters obtained from the fits shown in Figure~\ref{fig:laca33mn3}. A stretched  exponential $A_{rlx} \exp(-t/T_1)^K$ is linear with slope $(1 - K)$ on a plot of this type.}
\label{fig:laca33mn4}
\end{figure}

\begin{figure}
\caption{Time dependence of the $\mu$SR relaxing asymmetry $G_{rlx}$ (open squares) for zero applied field in La$_{0.67}$Ca$_{0.33}$MnO$_3$ showing the curves $\lambda_f$ (slowly relaxing curve corresponding to faster Mn spins) and $\lambda_s$ (rapidly relaxing curve corresponding to more slowly relaxing Mn spins) for a two exponential fit.  The curve for $\lambda_s$ has been offset from the data for clarity.}
\label{fig:asymm}
\end{figure}

\begin{figure}
\caption{Temperature dependence of the $\mu$SR fast ($\lambda_f$, top) and slow ($\lambda_s$, bottom) spin-lattice relaxation rates for zero applied field in La$_{0.82}$Ca$_{0.18}$MnO$_3$ obtained from a fit to Eq. ~(\ref{eq:strexp}) [closed symbols, exponent $K \simeq 1$ in Eq. ~(\ref{eq:strexp})] and Eq.~(\ref{eq:twoexp}) [open symbols].}
\label{fig:laca18mnfs}
\end{figure}

\begin{figure}
\caption{Temperature dependence of the $\mu$SR fast ($\lambda_f$, top) and slow ($\lambda_s$, bottom) spin-lattice relaxation rates for zero applied field in La$_{0.67}$Ca$_{0.33}$MnO$_3$ obtained from a fit to Eq. ~(\ref{eq:strexp}) [closed symbols, exponent $K \simeq 1$ in Eq. ~(\ref{eq:strexp})] and Eq.~(\ref{eq:twoexp}) [open symbols].}
\label{fig:laca33mnfs}
\end{figure}

\begin{figure}
\caption{Temperature dependence of the $\mu$SR fast ($A_f$, top) and slow ($A_s$, bottom) amplitudes for zero applied field in La$_{0.67}$Ca$_{0.33}$MnO$_3$ and La$_{0.82}$Ca$_{0.18}$MnO$_3$ obtained from a fit to Eq. ~(\ref{eq:strexp}) [closed symbols, exponent $K \simeq 1$ in Eq. ~(\ref{eq:strexp})] and Eq.~(\ref{eq:twoexp}) [open symbols].}
\label{fig:lacaamps}
\end{figure}


\begin{references}
\bibitem{note1}Note that if the $x = 0.33$ data is fit over the entire temperature range ($T = 10 - 267$ K) one obtains $T_C = 272 \pm 2$ K and $\beta = 0.32 \pm 0.02$, in agreement with previously published results\cite{Heffner96}.
\end{references}
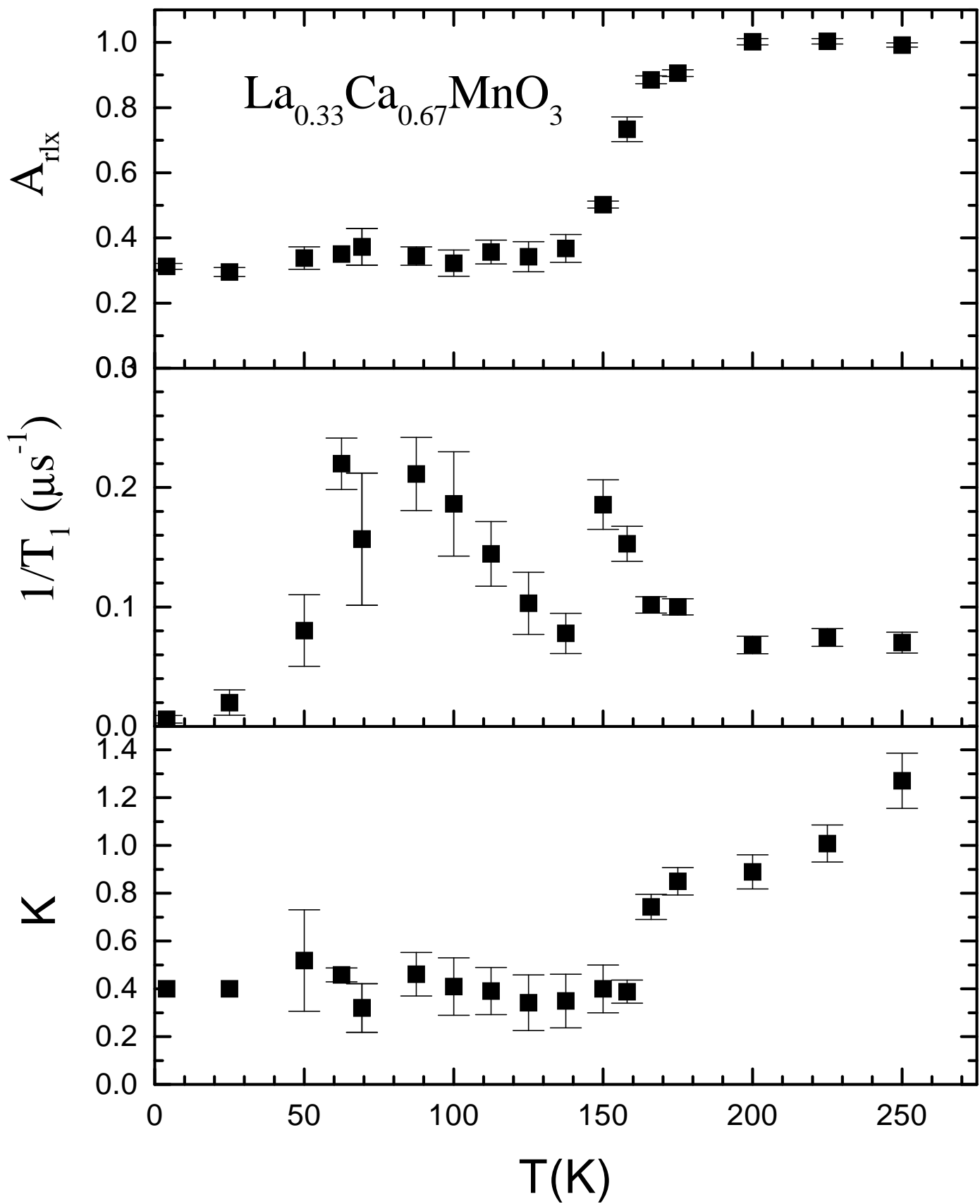
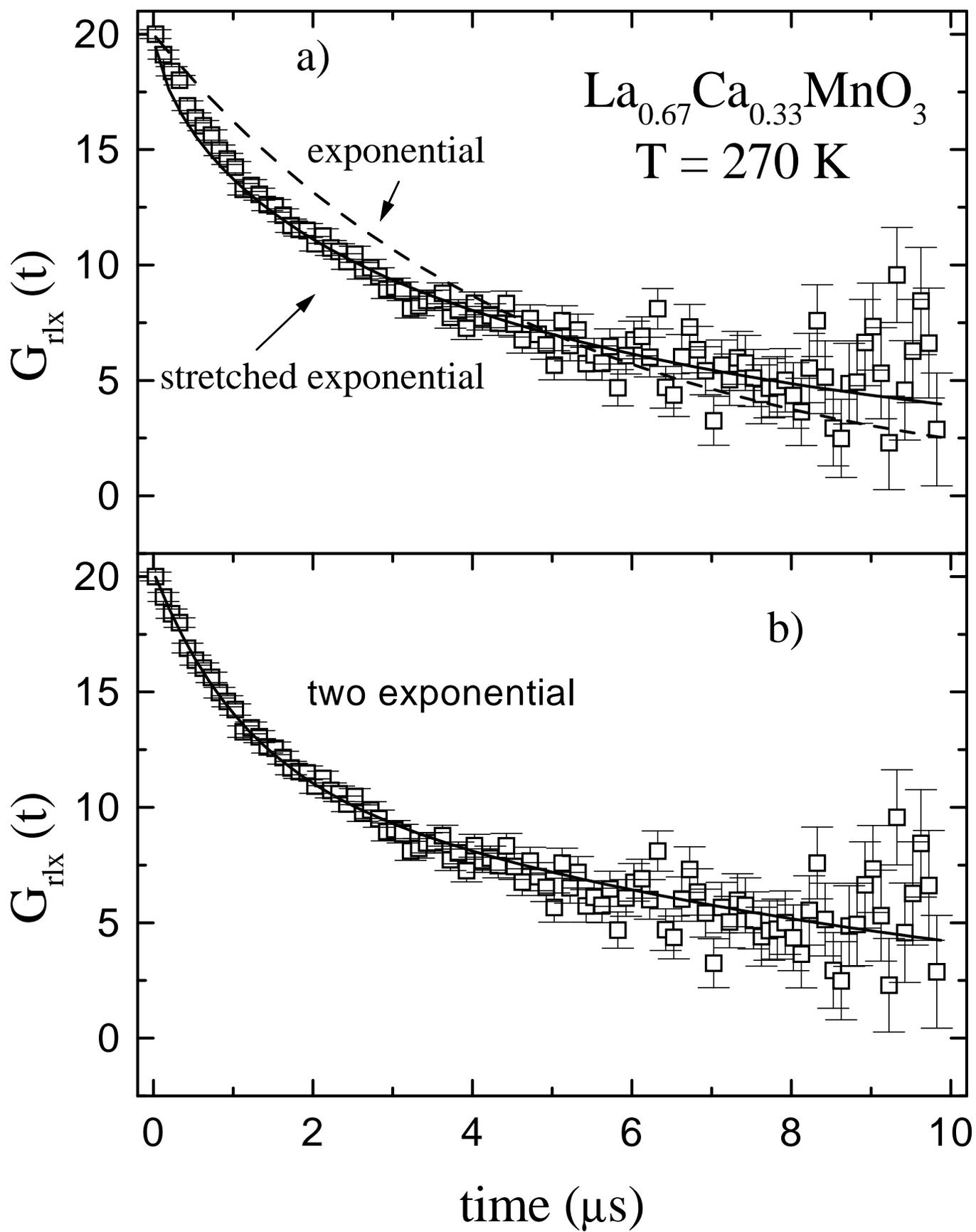
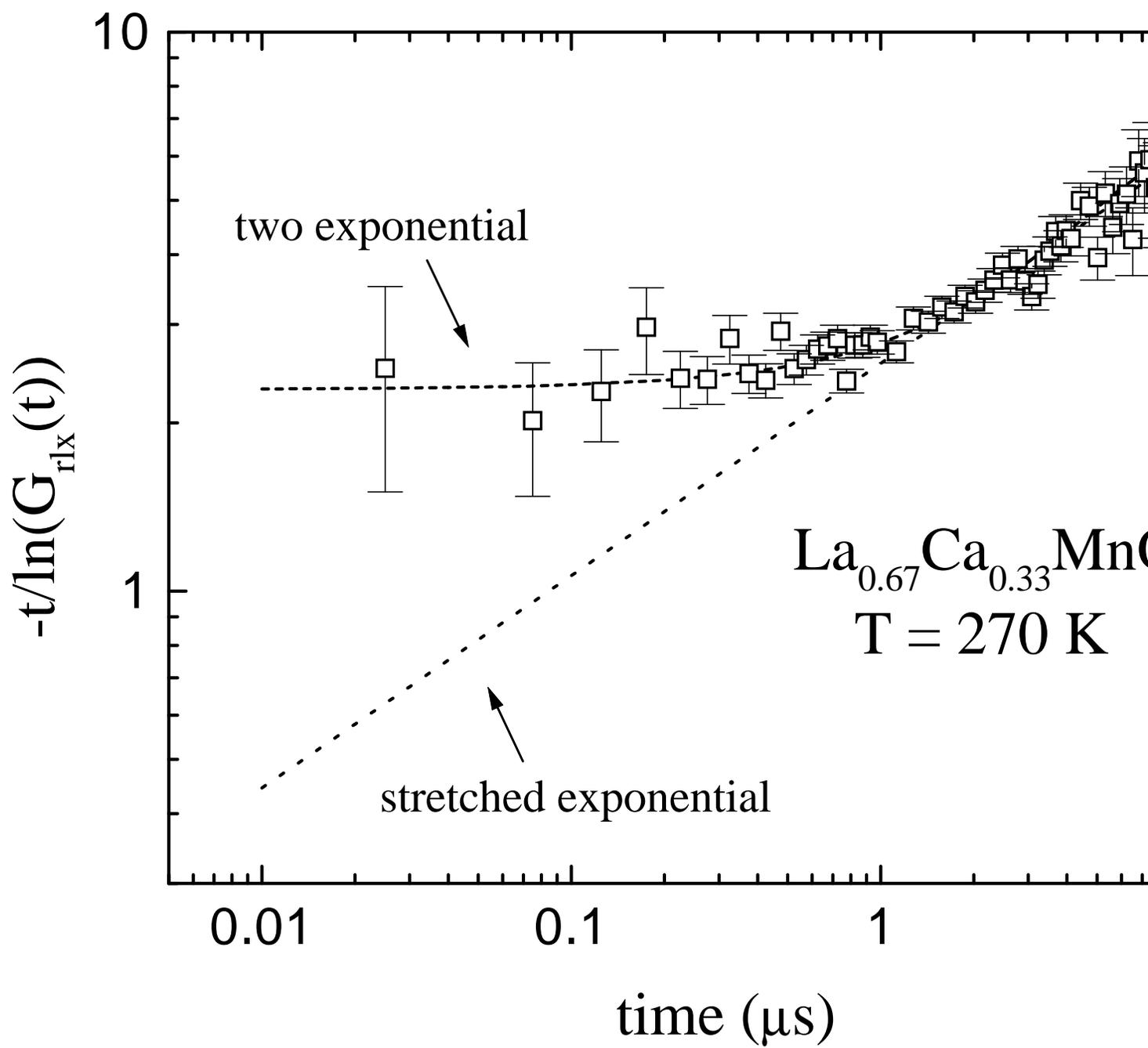
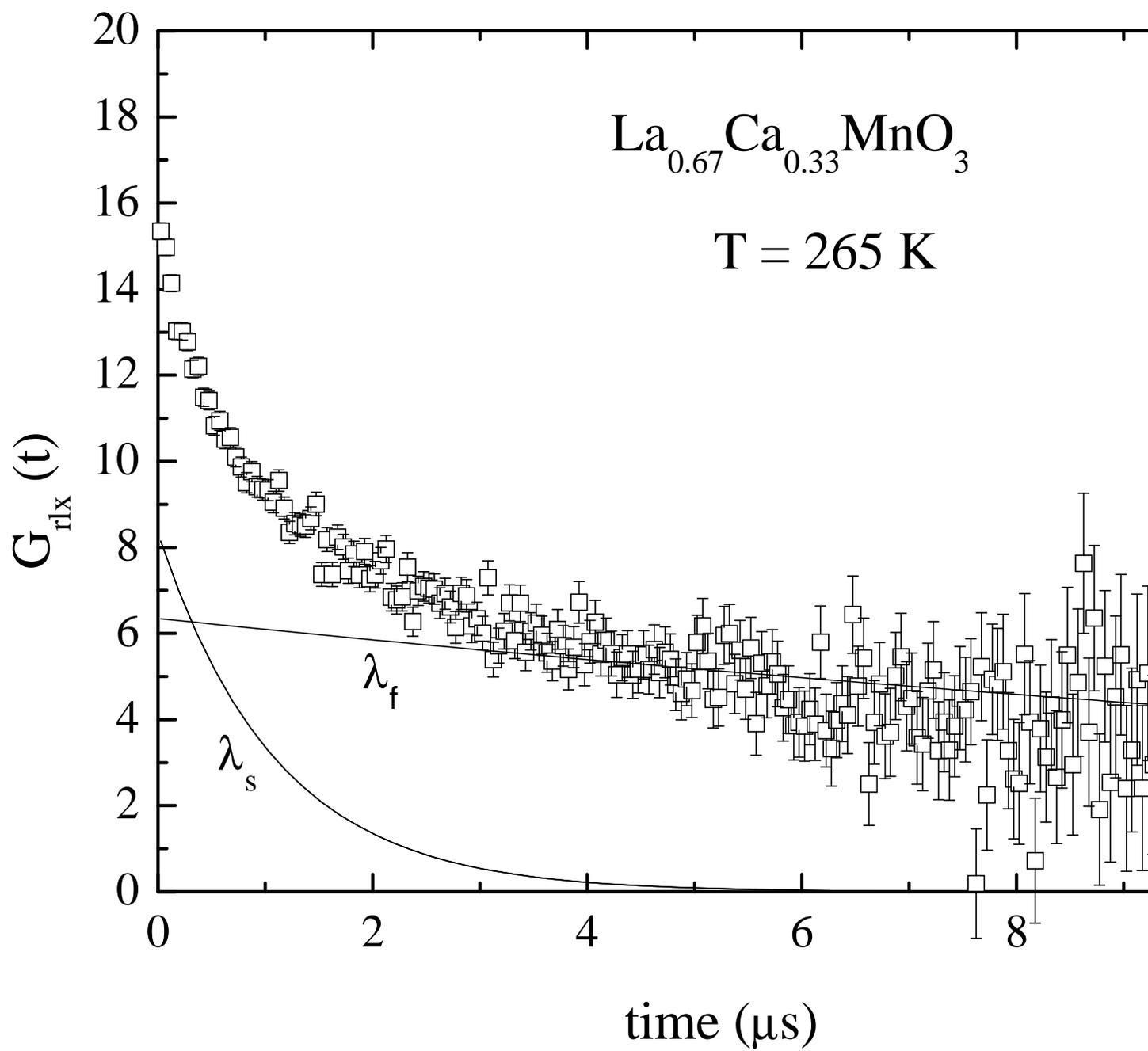
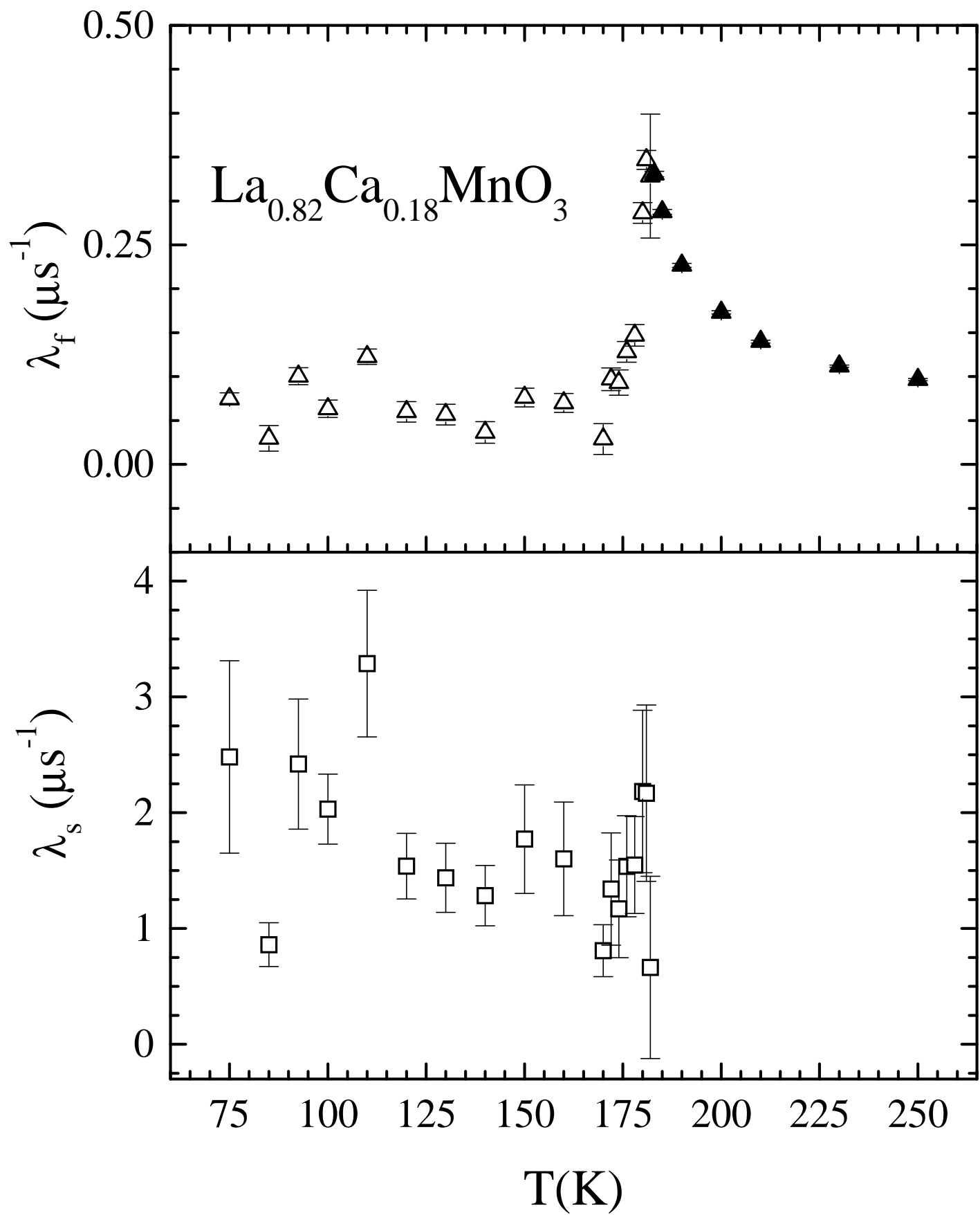
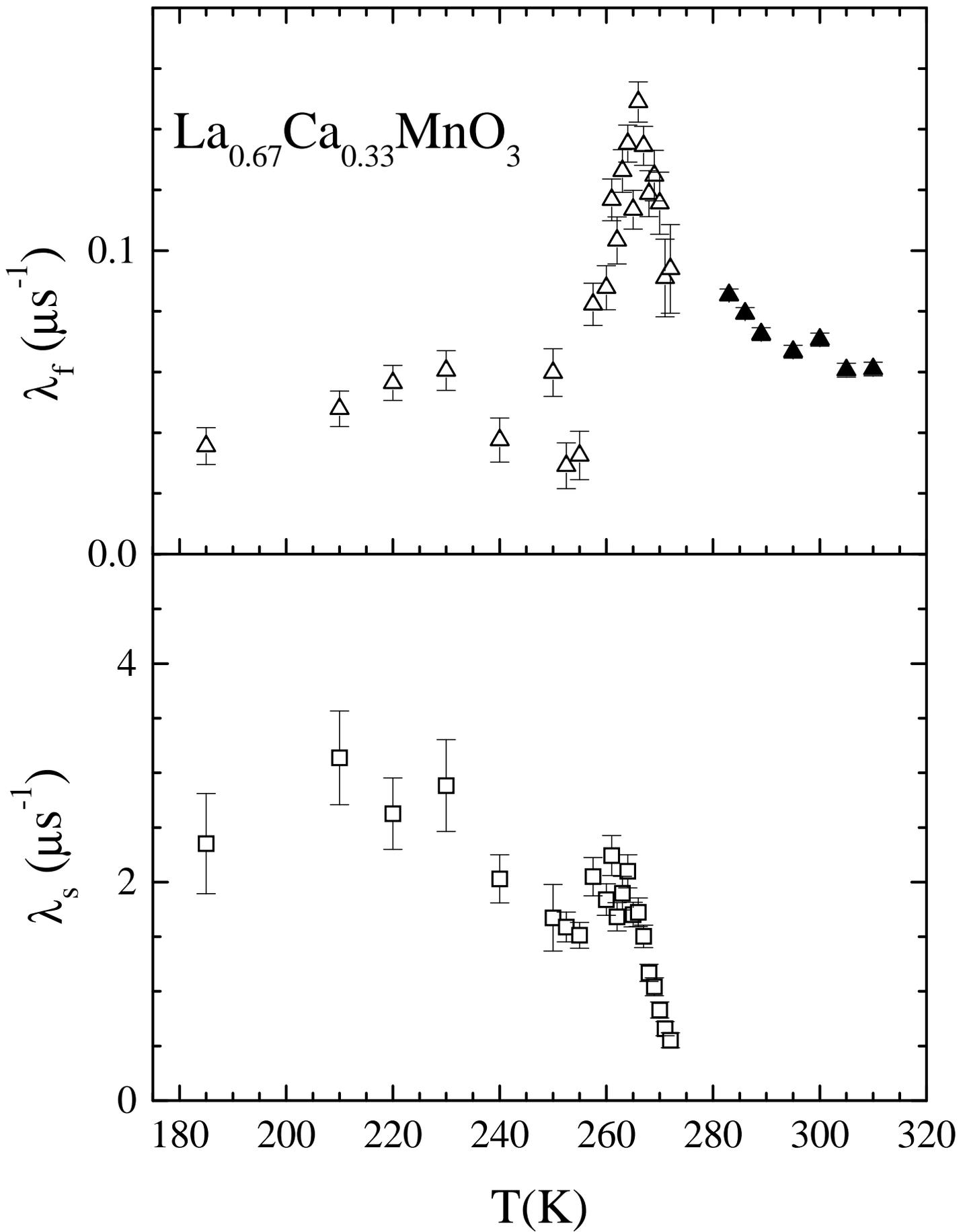
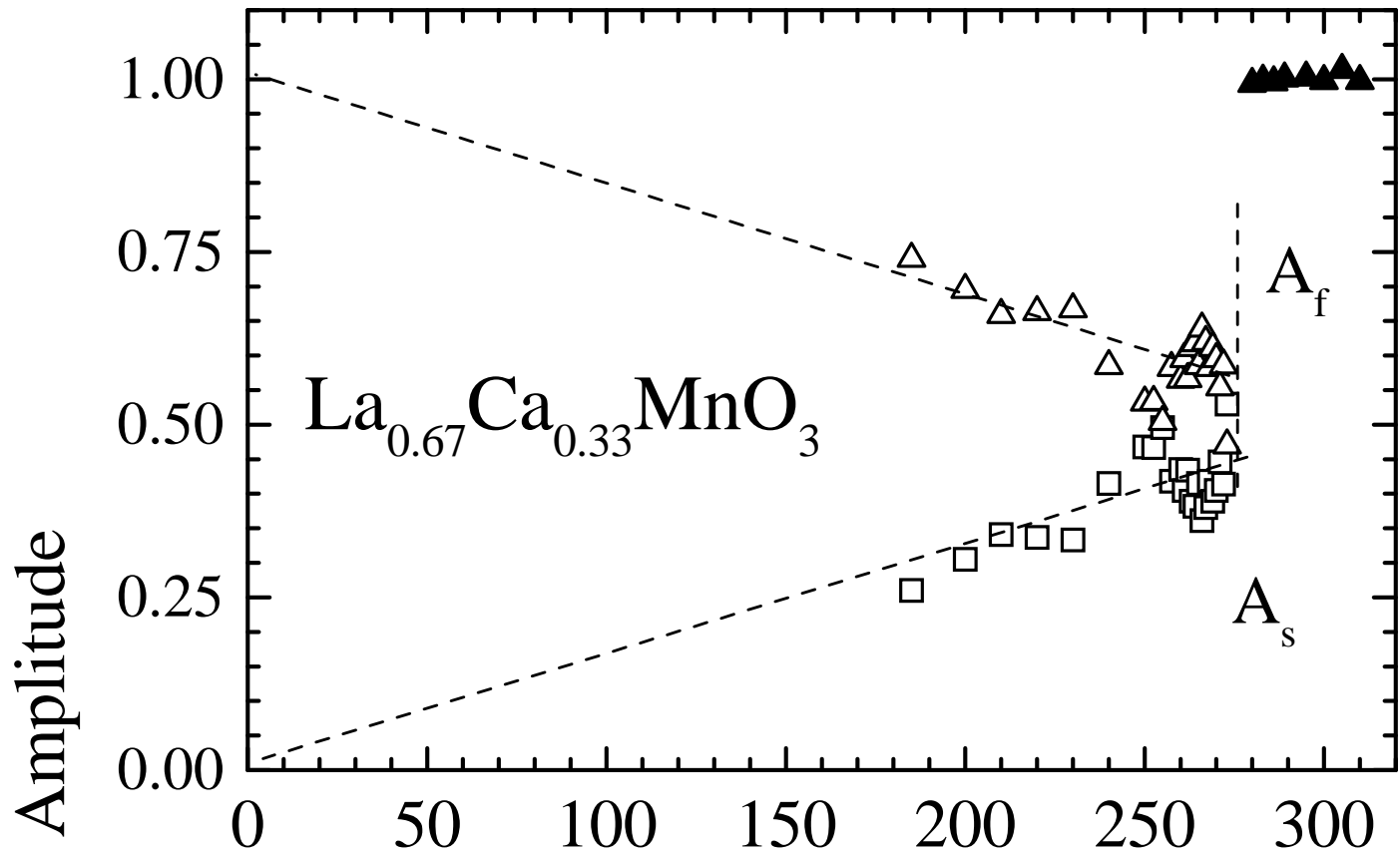
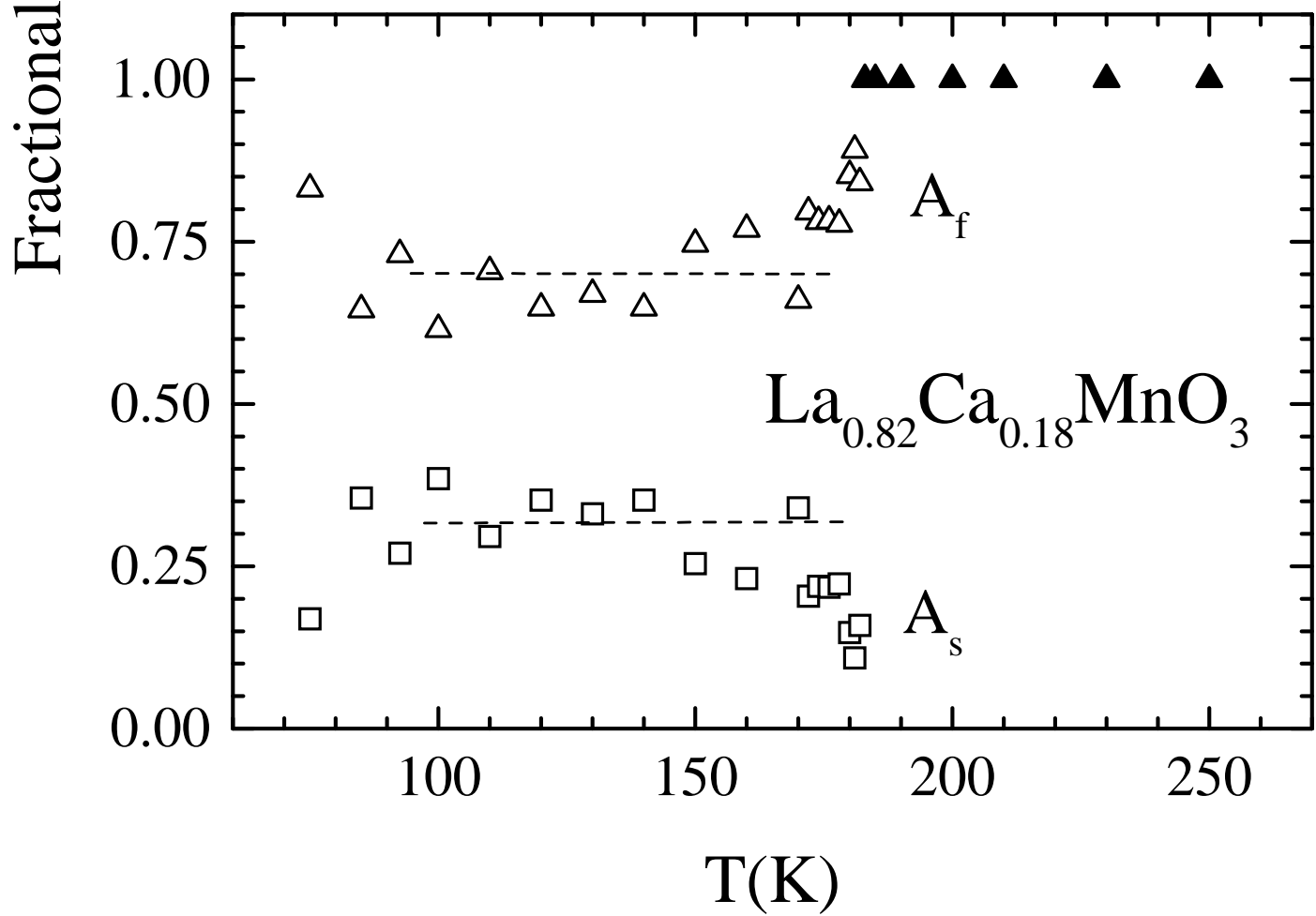
\end{document}